\documentclass[conference]{IEEEtran}
\usepackage{amsmath, amssymb, amsthm, amsfonts, amsxtra, latexsym, amscd}
\usepackage{amsthm}
\usepackage{mathtools}
\usepackage{amssymb}
\usepackage{lscape}
\usepackage{epsf}
\usepackage{color}
\usepackage{todonotes}

\newcommand{\comment}[1] {}

\newtheorem{theorem}{\indent Theorem}
\newtheorem{lemma}[theorem]{\indent Lemma}

\newtheorem{property}{\indent Property}
\newtheorem{example}{\indent Example}
\newtheorem{definition}{\indent Definition}

\newcommand{\code}{{\mathcal{C}}}

\newcommand{\distance}{{\mathsf{d}}}

\newcommand{\sH}{{\mathsf{H}}}

\newcommand{\weight}{{\mathsf{w}}}

\newcommand{\Code}{{\mathbb{C}}}

\newcommand{\ff}{{\mathbb{F}}}

\newcommand{\blde}{{\mbox{\boldmath $e$}}}
\newcommand{\bldf}{{\mbox{\boldmath $f$}}}

\newcommand{\bldG}{{\mbox{\boldmath $G$}}}

\newcommand{\bldM}{{\mbox{\boldmath $M$}}}

\newcommand{\bldx}{{\mbox{\boldmath $x$}}}
\newcommand{\bldy}{{\mbox{\boldmath $y$}}}
\newcommand{\bldz}{{\mbox{\boldmath $z$}}}

\newcommand{\bldzero}{{\mbox{\boldmath $0$}}}

\newcommand{\define}{\stackrel{\triangle}{=}}

\DeclarePairedDelimiter\Biggfloor{\Bigg\lfloor}{\Bigg\rfloor}

\begin{document}

\title{Linear Batch Codes}

\author{
    \IEEEauthorblockN{\bf Helger Lipmaa and Vitaly Skachek$^1$}
   \IEEEauthorblockA{Institute of Computer Science\\
    University of Tartu\\
    J. Liivi 2, Tartu 50409, Estonia\\
    Email: {\tt \{helger.lipmaa, vitaly.skachek\} @ut.ee}}
}
\maketitle

\footnotetext[1]{The work of the authors is supported in part by the research grants PUT405 and IUT2-1 from the Estonian Research Council and by the European Regional Development Fund through the Estonian Center of Excellence in Computer Science, EXCS. The work of V. Skachek is also supported in part by the EU COST Action IC1104.}  

\begin{abstract}
In an application, where a client wants to obtain many elements from a large database, it is often desirable to have some load balancing. Batch codes (introduced by Ishai \emph{et al.} in STOC 2004) make it possible to do exactly that: the large database is divided between many servers, so that the client has to only make a small number of queries to every server to obtain sufficient information to reconstruct all desired elements. Other important parameters of the batch codes are total storage and the number of servers. Batch codes also have applications in cryptography (namely, in the construction of multi-query computationally-private information retrieval protocols).

In this work, we initiate the study of \emph{linear} batch codes. These codes, in particular, are of potential use in distributed storage systems. 
We show that a generator matrix of a binary linear batch code is also a generator matrix of classical binary linear error-correcting code. 
This immediately yields that a variety of upper bounds, which were developed for error-correcting codes, are applicable also to
binary linear batch codes. We also propose new methods to construct large linear batch codes from the smaller ones. 
\end{abstract}


\section{Introduction}

Consider the scenario where a client wants to retrieve many (say $m$) elements from an $n$ element database, stored by a storage provider.
Accessing a single server by all clients simultaneously can create serious performance problems.
A simple solution is to duplicate the whole database between some $M$ servers, so that the client can query approximately $m / M$ elements from every server.
However, that solution is very costly storage-wise, since all servers together have then to store $N = M n$ database elements.

The things get even more problematic in the cryptographic scenario.
In an $m$-out-of-$n$ CPIR (computationally-private information retrieval~\cite{Kushilevitz}), the client wants to retrieve $m$ elements from an $n$ element database without the storage provider getting to know which elements were retrieved. An additional problem in this case is the storage provider's computational complexity that is $\Theta (n)$ per query in almost all known $1$-out-of-$n$ CPIR protocols. (The only exception is~\cite{Lipmaa}, where the per-query computational complexity is $O (n / \log n)$.) Just performing $m$ instances of an $1$-out-of-$n$ CPIR protocol would result in a highly prohibitive computational complexity.

To tackle both mentioned problems, Ishai \emph{et al.}~\cite{Ishai} proposed to use \emph{batch codes}.
More precisely, let $\Sigma$ be a finite alphabet.  
In an $(n, N, m, M, T)_{\Sigma}$ batch code, a database $\bldf$ of $n$ strings in $\Sigma$ is divided into $M$ buckets where each bucket contains $N  / M$ strings in $\Sigma$. (W.l.o.g., we assume that $M \mid N$.) If a client is to obtain $m$ elements of the original database, he query (no more than) $T$ elements from each of the $M$ buckets. A batch code guarantees that based on the answers to the resulting $\leq M \cdot T$ queries, the client is able to efficiently reconstruct the $m$ elements he was originally interested in. 

When using a batch code, the storage provider only needs to store $N$ elements. In addition, in the cryptographic scenario, the total computational complexity of the servers is $O (M T)$. Therefore, one is interested in simultaneously minimizing all three values $N$, $M$ and $T$.

Several different batch codes were proposed in~\cite{Ishai}.
Batch codes have been recently studied very actively in the combinatorial setting. Namely, a \emph{combinatorial batch code}~(CBC) satisfies the additional requirement that every element of every bucket is equal to some element of the original database. (See for example~\cite{Brualdi,Bhattacharya,Bujtas}.)
New constructions of combinatorial batch codes, based on affine planes and transversal designs, were recently presented in~\cite{Gal}.

CBCs suffer from some limitations. First, the requirement that each element in the bucket is equal to the element of the original database is generally not necessary. Relaxing this requirement can potentially lead to better parameter trade-offs. Second, batch codes are usually constructed using designs and related combinatorial structures. However, if such designs are employed in the practical settings, it might be difficult to come up with efficient retrieval algorithms for such codes due to the not-so-compact representation of the codes. As an alternative, we propose linear batch codes, where each bucket contains a linear combination of the elements in the original database. By using their flexible algebraic structure, better codes and more efficient retrieval algorithms can potentially be developed. 

We stress that linear batch codes are also well suitable for the use in the \emph{distributed data storage}~\cite{Dimakis}. 
The buckets can be viewed as servers. The reading of the requested data can be done ``locally'' from a small number of servers (for example, from those that are closer to the user, or connected by a higher-throughput links). The linear batch codes are generally fault-tolerable: if a small number of buckets stopped functioning, the data can be reproduced by reading data from (a small number) of other buckets. However, in order to achieve better \emph{locality} and \emph{repair bandwidth} (see~\cite{Dimakis}), linear batch codes with \emph{sparse} generator matrices can be particularly useful. 

In this paper, we develop a novel framework for analysis of linear batch codes, which is similar to that of classical error--correcting codes (ECCs).
In particular, the encoding is represented by multiplying an information vector by an analog of a generator matrix. As we show, generator matrices of good binary linear batch codes are also generator matrices of good classical ECCs. This immediately gives us a set of tools and bounds from the classical coding theory for analyzing binary linear batch codes. The converse, however, is not true: not every good binary linear ECC is a good linear batch code. Then, we present a number of simple constructions of larger linear batch codes from the smaller ones. It worth mentioning that this novel framework opens a number of research directions related to linear batch codes. We conclude the paper with the list of some of the open questions. 

The paper is structured as follows. The notations and some known results are presented in Section~\ref{sec:notation}. The properties of linear batch codes are analyzed in Section~\ref{sec:linear}. New construction methods of linear batch codes are presented in Section~\ref{sec:constructions}. The paper is summarized in Section~\ref{sec:discussion}.

\section{Notation and known results}
\label{sec:notation}

Let $[n] \triangleq \{1, 2, \cdots, n \}$. 
We use notation $\distance_H(\bldx, \bldy)$ to denote the Hamming distance between the vectors 
$\bldx$ and $\bldy$, and notation $\weight_H(\bldx)$ to denote the Hamming weight of $\bldx$. 
We also denote by $\bldzero$ the row vector consisting of all zeros, and by $\blde_i$ the row
vector having one at position $i$ and zeros elsewhere (the length of vectors will be clear from the context). 
The binary entropy function is defined as $\sH_2(x) \triangleq - x \log_2 x - (1-x) \log_2 (1-x)$. 

\begin{definition}[\cite{Ishai}]
Let $\Sigma$ be a finite alphabet. We say that $\code$ is an $(n, N, m, M, t)_\Sigma$ batch code over a finite alphabet $\Sigma$ if 
it encodes any string $\bldx = (x_1, x_2, \cdots, x_n) \in \Sigma^n$ into $M$ strings (buckets) of total length $N$ 
over $\Sigma$, namely $\bldy_1, \bldy_2, \cdots, \bldy_M$,  such that for each $m$-tuple (batch) of (not neccessarily distinct)
indices $i_1, i_2, \cdots, i_m \in [n]$, the entries
$x_{i_1}, x_{i_2}, \cdots, x_{i_m}$ can be retrieved by reading at most $t$ symbols from each bucket.
The ratio $R \define n/N$ is called the rate of the code.
\end{definition}

If for the code $\code$ it holds that $t=1$, then we use notation $(n, N, m, M)_\Sigma$ for it. This corresponds to an important special case when 
only one symbol is read from each bucket. 

Note that the buckets in this definition correspond to the devices in the above example,
the encoding length $N$ to the total storage, and the parameter $t$ to the maximal load.

If $\Sigma = \ff_q$ is a finite field, we also use notation $(n, N, m, M, t)_q$ (or $(n, N, m, M)_q$) to denote $(n, N, m, M, t)_{\Sigma}$
(or $(n, N, m, M)_{\Sigma}$, respectively). 

\begin{definition}
We say that an $(n, N, m, M, t)_{q}$ batch code is \emph{linear}, if every entry of every bucket is a linear combination of original database elements. 
\end{definition}

Before going further, we recall the following code composition from~\cite[Lemma~3.5]{Ishai}. 
\begin{lemma}[Batch code composition]\label{fact:comp}
Let $\code_1$ be an $(n_1, N_1 = M_1 n_2, m_1, M_1)_{\Sigma}$ batch code and $\code_2$ an $(n_2, N_2, m_2, M_2)_{\Sigma}$ batch code such that the length of each bucket in $\code_1$ is $n_2$ (in particular, $N_1 = M_1 n_2$). Then, there exists an $(n = n_1, N = M_1 N_2, m = m_1 m_2, M = M_1 M_2)_{\Sigma}$ batch code $\code_1 \otimes \code_2$.
\end{lemma}
Thus, one can design batch codes by first considering special cases (like $M = N$), and then combining suitable batch codes to get rid of such restrictions.


\section{Linear batch codes}
\label{sec:linear}

We start with the following example, which is based on so-called ``subcube codes'' in~\cite[Section 3.2]{Ishai}.  
\begin{example}
Consider a database of $n$ elements over $\ff_q$, where the user wants to retrieve any $m$ of them. Let $t$ and $n$ be integers, $2 | n$. Divide the database 
$\bldx = (x_1, x_2, \cdots, x_{n})$ into two buckets, where bucket $i$, $i=1,2$, contains elements 
$(x_{(i-1)n/2+1}, x_{(i-1)n/2+2}, \cdots, x_{i \cdot n/2})$. The third bucket will contain elements 
$(x_{1}+x_{n/2+1}, x_2 + x_{n/2+2}, \cdots, x_{n/2} + x_{n})$. 

This code is a linear $[n, N = 1.5n, m = 2t, M = 3, t]_q$ code for any $1 \le t \le n/2$. Observe, however, that the proposed code can be 
viewed as $n/2$ copies of the same  $[2, 3, 2, 3, 1]_q$ subcube code. 
\end{example}

In what follows, we consider the case of a linear batch code $\code$ with 
$t=1$. Moreover, we limit ourselves to the case when $N = M$, 
which means that each encoded bucket contains just one element in $\ff_q$. 

\begin{definition}
For simplicity we refer to a linear $(n, N=M, m, M)_{q}$ batch code as $[M, n, m]_q$ batch code. 
\end{definition} 

As before, let $\bldx = (x_1, x_2, \cdots, x_n)$ be an information string, and let
$\bldy = (y_1, y_2, \cdots, y_M)$ be an encoding of $\bldx$. Due to linearity of the code, each encoded symbol 
$y_i$, $i \in [M]$, can be written as $y_i = \sum_{j=1}^n g_{j,i} x_j$ for some elements 
$g_{j,i} \in \ff_q$, $j \in [n]$, $i \in [M]$. Then we can form the matrix $\bldG$ as follows: 
\[
\bldG = \Big( g_{j,i} \Big)_{j \in [n], i \in [M]} \; ,
\]
and thus
\[
\bldy = \bldx \bldG \; . 
\]
The $n \times M$ binary matrix $\bldG$ play a role similar to generator matrix for a classical linear ECC. 
In the sequel, we will call $\bldG$ \emph{generator matrix} of the batch code $\code$. We denote by $\bldG_i$ the 
$i$-th row of $\bldG$ and by $\bldG^{[i]}$ the $i$-th column of $\bldG$. 
\medskip 

Observe that we can retrieve $x_j$ from $\bldy$ (for some $j \in [n]$) using $[M, n, m]_q$ batch code if there exists 
a linear combinations of columns in $\bldG$ over $\ff_q$, which is equal to $\blde_j$. 
Moreover, the following generalization of this property holds.  
\begin{property}
Let $\code$ be an $[M, n, m]_q$ batch code. It is possible to retrieve $x_{i_1}, x_{i_2}, \cdots, x_{i_m}$ simultaneously if there
exist $m$ non-intersecting sets of indices of columns in $\bldG$, and for the $r$-th set there exists a linear combination of columns of $\bldG$ indexed by that set, which equals to the column vector $\blde_{i_r}^T$, for all $r \in [m]$.
\label{prop:1}
\end{property}
\begin{proof}
Let 
\[
\bldG \triangleq \left[ \bldG^{[1]} \; | \;  \bldG^{[2]} \; | \; \cdots \; | \; \bldG^{[M]} \right] \; ,  
\]
where $\bldG^{[\ell]}$ is the $\ell$-th column in $\bldG$. 
Let $T_1, T_2, \cdots, T_m$ be non-intersecting sets of indices, such that for each $r \in [m]$ 
\[
\blde_{i_r}^T = \sum_{\ell \in T_r} \alpha_\ell \cdot \bldG^{[\ell]} \; , 
\]
where all $\alpha_\ell \in \ff_q$. 
Due to linearity, the encoding of $\bldx = (x_1, x_2, \cdots, x_n)$ can be written as 
\[
\bldy = (y_1, y_2, \cdots, y_M) = \bldx \cdot \bldG \; . 
\]
Then, 
\begin{eqnarray*}
x_{i_r} & = & \bldx \cdot \blde_{i_r}^T \\
& = & \bldx \cdot \left( \sum_{\ell \in T_r} \alpha_\ell \cdot \bldG^{[\ell]} \right) \\
& = & \sum_{\ell \in T_r} \alpha_\ell (\bldx \cdot \bldG^{[\ell]}) \\
& = & \sum_{\ell \in T_r} \alpha_\ell \cdot y_\ell \; , 
\end{eqnarray*}
and therefore the value of $x_{i_r}$ can be obtained by querring only the values of $y_\ell$ for $\ell \in T_r$. 
The conclusion follows from the fact that all $T_r$ do not intersect. 
\end{proof}
\medskip 

In the rest of the paper we assume that the retrieving server performs only linear operations over the columns of the matrix $\bldG$ (in other words, 
it only adds and subtracts $y_1$, $y_2$, $\cdots$, $y_M$, and multiplies them by the elements in $\ff_q$). This is a standard assumption in many areas of linear coding (in particular, in network and index coding). In that case, the condition in Property~\ref{prop:1} becomes both neccessary and sufficient.  

\begin{example}
Consider the following linear binary batch code $\code$ whose $4 \times 9$ generator matrix is given by 
\[
\bldG = \left( 
\begin{array}{ccccccccc}
1 & 0 & 1 & 0 & 0 & 0 & 1 & 0 & 1 \\
0 & 1 & 1 & 0 & 0 & 0 & 0 & 1 & 1 \\
0 & 0 & 0 & 1 & 0 & 1 & 1 & 0 & 1 \\  
0 & 0 & 0 & 0 & 1 & 1 & 0 & 1 & 1 
\end{array}
\right) \; . 
\]

Let $\bldx = (x_1, x_2, x_3, x_4)$, $\bldy = \bldx \bldG$. 

Assume that we want to retrieve the values of $(x_1, x_1, x_2, x_2)$. 
Consider, for example, the following combinations of the columns of $\bldG$: 
\begin{eqnarray*}
&& \left( 
\begin{array}{c}
1 \\
0 \\
0 \\
0
\end{array}
\right) \; , 
\left( 
\begin{array}{c}
1 \\
0 \\
0 \\
0
\end{array}
\right) = 
\left( 
\begin{array}{c}
0 \\
1 \\
0 \\
0
\end{array}
\right) + 
\left( 
\begin{array}{c}
1 \\
1 \\
0 \\
0
\end{array}
\right)
\; , \\
&& \left( 
\begin{array}{c}
0 \\
1 \\
0 \\
0
\end{array}
\right) 
= 
\left( 
\begin{array}{c}
0 \\
0 \\
0 \\
1
\end{array}
\right) 
+ \left( 
\begin{array}{c}
0 \\
1 \\
0 \\
1
\end{array}
\right) 
\; , \\
&& \left( 
\begin{array}{c}
0 \\
1 \\
0 \\
0
\end{array}
\right) = 
\left( 
\begin{array}{c}
0 \\
0 \\
1 \\
0
\end{array}
\right) + \left( 
\begin{array}{c}
0 \\
0 \\
1 \\
1 
\end{array}
\right) + 
\left( 
\begin{array}{c}
1 \\
0 \\
1 \\
0
\end{array}
\right) 
+
\left( 
\begin{array}{c}
1 \\
1 \\
1 \\
1
\end{array}
\right) 
\; .
\end{eqnarray*}
Then, we can retrieve $(x_1, x_1, x_2, x_2)$ from the following set of equations: 
\[
\left\{ 
\begin{array}{ccl}
x_1 & = & y_1 \\
x_1 & = & y_2 + y_3 \\
x_2 & = & y_5 + y_8 \\
x_2 & = & y_4 + y_6 + y_7 + y_9 
\end{array} \right. \; . 
\]
Moreover, it is straightforward to verify that any $4$-tuple $(x_{i_1}, x_{i_2}, x_{i_3}, x_{i_4})$, where $i_1, i_2, i_3, i_4 \in [4]$, can be retrieved by using columns indexed by some four non-intersecting sets of indices in $[9]$. 
Therefore, the code $\code$ is a $[9, 4, 4]_2$ batch code. As a matter of fact, this code is the two-layer construction of ``subcube code'' in~\cite[Section 3.2]{Ishai}.  

\end{example}

Next, we state the following simple lemmas. 

\begin{lemma}
Let $\code$ be an $[M, n, m]_q$ batch code. Then, each row of $\bldG$ has Hamming weight at least $m$. 
\label{lemma:distance}
\end{lemma} 

{\bf Proof.} Consider row $j$, for an arbitrary $j \in [n]$. We can retrieve the combination $(x_j, x_j, \cdots, x_j)$ if there are $m$ non-intersecting sets of columns, such that sum of the elements in each set is equal $\blde_j^T$. Therefore, there are at least $m$ columns in $\bldG$ with a nonzero entry in position $j$.
\qed

\begin{lemma}
Let $\code$ be an $[M, n, m]_q$ batch code. Then, the matrix $\bldG$ is a full rank matrix. 
\label{lemma:full-rank}
\end{lemma} 

{\bf Proof.} We should be able to recover any combination of size $m$ of $\{ x_1, x_2, \cdots, x_n \}$.
Then, the column vectors 
\begin{multline*}
\left( 
\begin{array}{c}
1 \\
0 \\
0 \\
\vdots \\
0
\end{array}
\right) , \; 
\left( 
\begin{array}{c}
0 \\
1 \\
0 \\
\vdots \\
0
\end{array}
\right) , \; 
\left( 
\begin{array}{c}
0 \\
0 \\
1 \\
\vdots \\
0
\end{array}
\right) , \quad \cdots \quad   
\left( 
\begin{array}{c}
0 \\
0 \\
0 \\
\vdots \\
1
\end{array}
\right)
\end{multline*}
are all in the column space of $\bldG$. Therefore, the column space of $\bldG$ has dimension $n$, and so the
matrix is full rank. \qed
\medskip

The following theorem is the main result of this section. The presented proof of this theorem works only for \emph{binary} batch codes. 
However, binary codes are very important special case of batch codes, as typical practical applications use
binary representation of information. The proof uses the fact that the codes are binary, --- we are not aware of a simple generalization of this proof to nonbinary case. 
\begin{theorem}
Let $\code$ be an $[M, n, m]_2$ batch code $\code$ over $\ff_2$. Then, $\bldG$ is a generator matrix 
of the classical error-correcting $[M, n, \ge m]_2$ code. 
\label{thrm:batch=linear}
\end{theorem}

{\bf Proof.} 
Let $\Code$ be a classical ECC, whose generating matrix is $\bldG$. 
It is obvious that the length of $\Code$ is $M$. Moreover, since the matrix $\bldG$ 
is a full rank matrix due to Lemma~\ref{lemma:full-rank}, we obtain that the dimension of $\Code$ is $n$. 
Thus, the only parameter in question is the minimum distance of $\Code$. 

In order to show that the minimum distance of $\Code$ is at least $m$, it will be sufficient to 
show that any non-zero linear combination of the rows of $\bldG$ has Hamming weight at least $m$. 
Consider an arbitrary linear combination of the rows of $\bldG$, whose indices are given by a set $T \neq \varnothing$,  
\[
\bldz = \sum_{i \in T} \bldG_i \; . 
\]
Take an arbitrary index $i_0 \in T$. Due to the properties of the batch codes we should 
be able to recover $(x_{i_0}, x_{i_0}, \cdots, x_{i_0})$ from $\bldy$. Therefore, there 
exist $m$ disjoint sets of indices $S_1, S_2, \cdots, S_m$, $S_i \subseteq [M]$, such that 
for all $i \in [m]$: 
\begin{equation}
\sum_{j \in S_i} \bldG^{[j]} = \blde_{i_0}^T \;  .  
\label{eq:unity}
\end{equation}

Now, consider the sub-matrix $\bldM_i$ of $\bldG$ which is formed by the rows of $\bldG$ indexed by $T$ 
and the columns of $\bldG$ indexed by $S_i$. Due to~(\ref{eq:unity}), the row of $\bldM_i$ that corresponds 
to the row $i_0$ in $\bldG$, has an odd number of ones in it. All other rows of $\bldM_i$ contain an even number of ones. 
Therefore, the matrix $\bldM_i$ contains an odd number of ones. This means that the vector 
of $\bldz$ will also contain an odd number of ones in the positions given by the set $S_i$.  
This odd number is at least one. 

We conclude that $\bldz$ contains at least one `$1$' in positions given by the set $S_i$, for all $i \in [m]$. 
The sets $S_i$ are disjoint, and therefore the Hamming weight of $\bldz$ is at least $m$. 
\qed

\begin{example}
The converse of Theorem~\ref{thrm:batch=linear} is generally not true. 
In other words, if $\bldG$ is a generator matrix 
of a classical error-correcting $[M, n, m]_2$ code, then 
the corresponding code $\code$ is not necessarily an $[M, n, m]_2$ batch code. 
For example, take $\bldG$ to be a generator matrix of the classical $[4, 3, 2]_2$ ECC as follows: 
\[
\bldG = \left( \begin{array}{cccc}
1 & 1 & 1 & 1 \\
0 & 1 & 0 & 1 \\
0 & 0 & 1 & 1 \\
\end{array}
\right) \; . 
\]
Let $\bldx = (x_1, x_2, x_3)$, $\bldy = (y_1, y_2, y_3, y_4) = \bldx \bldG$. 

It is impossible to retrieve $(x_2, x_3)$. This can be verified by the fact that 
\[
x_2 = y_1 + y_2 = y_3 + y_4 \quad \mbox{and} \quad  x_3 = y_1 + y_3 = y_2 + y_4 \; ,
\]
and so one of the $y_i$'s is always needed to compute each of $x_2$ and $x_3$. 
\end{example}

{\bf Corollary.} The topic of linear ECCs was very intensively studied over the years. 
Various well-studied properties of linear ECCs, such as MacWilliams identities~\cite{MacWilliams}, apply also 
to linear batch codes due to Theorem~\ref{thrm:batch=linear} (for $t=1$, $M=N$ and $q=2$).
A variety of bounds on the parameters of ECCs, such as sphere-packing bound~(\ref{eq:spacking-bound}), Plotkin bound~(\ref{eq:plotkin-bound}), 
Griesmer bound~(\ref{eq:griesmer-bound}), Elias-Bassalygo bound~(\ref{eq:elias-bound}), 
McEliece-Rodemich-Rumsey-Welch bound~(\ref{eq:MRRW-bound})~\cite{Rodemich} (see also~\cite[Chapter 4]{Roth-book},~\cite{MacWilliams-Sloane})
apply to the parameters of linear binary $[M, n, m]$ batch codes.
 
\begin{equation}
2^{M-n} \ge \sum_{i=0}^{\lfloor (m-1)/2 \rfloor} {M \choose i} \; 
\label{eq:spacking-bound}
\end{equation}
\begin{equation}
m \le \frac{M \cdot 2^{n-1}}{2^{n}-1} \; 
\label{eq:plotkin-bound}
\end{equation}
\begin{equation}
M \ge \sum_{i=0}^{n-1} \Big\lceil \frac{m}{2^i} \Big\rceil \; 
\label{eq:griesmer-bound}
\end{equation}
\begin{equation}
\frac{n}{M} \le 1 - \sH_2 \left( \frac{1}{2} \left( 1 - \sqrt{1 - 2 \frac{m}{M}} \right) \right) + o(1) \; 
\label{eq:elias-bound}
\end{equation}
\begin{equation}
\frac{n}{M} \le \sH_2 \left( \frac{1}{2} - \frac{\sqrt{m(M-m)}}{M} \right)  + o(1) \; 
\label{eq:MRRW-bound}
\end{equation}

\section{Constructions of New Codes}
\label{sec:constructions}

In this section we present several simple methods to construct new linear batch codes from the existing ones. 

\begin{theorem}
Let $\code_1$ be an $[M_1, n, m_1]_q$ batch code and $\code_2$ be an $[M_2, n, m_2]_q$ batch code. 
Then, there exists an $[M_1+ M_2, n, m_1 + m_2]_q$ batch code. 
\end{theorem}

{\bf Proof.} 
Let $\bldG_1$ and $\bldG_2$ be $n \times M_1$ and $n \times M_2$ generator 
matrices corresponding to $\code_1$ and $\code_2$, respectively. 
Consider the following $n \times (M_1+M_2)$ matrix 
\[
\hat{\bldG} = \left[ \; \bldG_1 \; | \; \bldG_2 \; \right] \; . 
\]
This matrix corresponds to a batch code of length $M_1 + M_2$ with $n$ variables. 
It is sufficient to show that any combination of $m_1+m_2$ variables can be retrieved. 
By the assumption, the first (any) $m_1$ variables can be retrieved from the first $M_1$ coordinates 
of $\bldy$ and the last $m_2$ variables can be retrieved from the last $M_2$ coordinates 
of $\bldy$. This completes the proof. 
\qed

\begin{theorem}
Let $\code_1$ be an $[M_1, n_1, m_1]_q$ batch code and $\code_2$ be an $[M_2, n_2, m_2]_q$ batch code. 
Then, there exists an $[M_1+ M_2, n_1 + n_2, \min\{m_1, m_2\}]_q$ batch code. 
\end{theorem}

{\bf Proof.} 
As before, denote by $\bldG_1$ and $\bldG_2$ the $n_1 \times M_1$ and $n_2 \times M_2$ generator 
matrices corresponding to $\code_1$ and $\code_2$, respectively. 
Consider the following $(n_1 + n_2) \times (M_1+M_2)$ matrix 
\[
\hat{\bldG} = \left[ \begin{array}{c|c}
 \bldG_1 & \bldzero \\
 \hline
 \bldzero & \bldG_2 
\end{array} \right] \; . 
\]
The matrix $\hat{\bldG}$ corresponds to a batch code of length $M_1 + M_2$ with $n_1 + n_2$ variables. 
Moreover, any combination of $\min\{m_1, m_2\}$ variables can be retrieved. 
If all unknowns are from $\{x_1, x_2, \cdots, x_{n_1} \}$, then they can be retrieved by
using only the first $M_1$ columns of $\hat{\bldG}$. If all unknowns are from $\{ x_{n_1+ 1}, x_{n_1 + 2}, \cdots, x_{n_1+n_2} \}$, then they can be retrieved by using only the last $M_2$ columns of $\hat{\bldG}$. 
Generally, some unknowns can be retrieved by using combinations of the first $M_1$ columns, while the other unknowns 
are retrieved using combinations of the last $M_2$ columns. Since the number of unknowns is at most $\min\{m_1, m_2\}$, 
we can always retrieve all of them simultaneously. 
\qed

The next theorem presents another construction of batch code from a smaller batch code. 

\begin{theorem}
Let $\code$ be an $[M, n, m]_q$ batch code, and let $\bldG$ be the corresponding $n \times M$ matrix. 
Then, the code $\hat{\code}$, defined by the $(n + 1) \times (M + m)$ matrix 
\begin{eqnarray*}
& \hat{\bldG} = \left( \begin{array}{ccccc|cccc} 
&   &   &   &        &  0 & 0 & \cdots & 0  \\
&   &  \bldG &    &        &  0 & 0 & \cdots & 0  \\
&   &   &   &        &  \vdots & \vdots & \cdots & \vdots  \\
&   &   &   &        &  0 & 0 & \cdots & 0  \\
\hline 
\bullet & \bullet & \bullet & \cdots & \bullet & 1 & 1 & \cdots & 1 \\
\end{array}
\right) \\
& \phantom{ooo} \underbrace{\phantom{oooooooooooooooooi}}_{M} \underbrace{\phantom{ooooooooooooo}}_{m} 
\end{eqnarray*} 
is an $[M+m, n+1, m]$ batch code, where $\bullet$ stands for an arbitrary element in $\ff_q$.
\end{theorem}

{\bf Proof} As before, let $\bldx = (x_1, x_2, \cdots, x_n, x_{n+1})$ and $\bldy = (y_1, y_2, \cdots, y_{M+m}) = \bldx \hat{\bldG}$. Assume that we want to retrieve the vector $\bldz = (x_{i_1}, x_{i_2}, \cdots, x_{i_m})$. 

Take a particular $x_{i_j}$ in $\bldz$, $j \in [m]$. Consider two cases. If ${i_{j}} \neq {n+1}$ then, since $\code$ is a batch code, 
we have 
\[
x_{i_j} = \sum_{\ell \in T_{i_j}} y_\ell \; + \; \xi \cdot x_{n+1} \; ,  
\]
where $T_{i_j} \subseteq [M]$ and $\xi \in \ff_q$. In that case, if $\xi = 0$, then $x_{i_j} = \sum_{\ell \in T_{i_j}} y_\ell$. 
If $\xi \neq 0$, then $x_{i_j} = \sum_{\ell \in T_{i_j}} y_\ell + \xi \cdot y_{M+j}$. Observe that all $T_{i_j}$ are disjoint due to 
the properties of a batch code. 

In the second case, ${i_{j}} = {n+1}$, and we simply set $x_{i_j} = x_{n+1}= y_{M+j}$. 

In both cases, we used sets $\{ y_\ell \; : \; \ell \in T_{i_j} \cup \{M + j\} \}$ to retrieve $x_{i_j}$. These sets are all disjoint for $j \in [m]$. 

We conclude that all $m$ unknowns $x_{i_j}$, $j \in [m]$, can be retrieved simultaneously. 
\qed

\section{Discussion}
\label{sec:discussion}

In this paper, we studied \emph{linear} batch codes. We defined generator matrices of such codes. We also showed that a generator matrix of 
a linear $[M, n, m]_2$ batch code is also a generator matrix of a classical $[M, n, m]_2$ ECC. The converse is 
not neccessarily true. Finally, we presented several simple ways to construct new linear batch codes from smaller codes. 

Since linear batch codes are closely related to linear ECCs, 
various well-known properties of linear ECCs, and in particular bounds on their parameters~(\ref{eq:spacking-bound})-(\ref{eq:MRRW-bound}) apply also 
to linear binary batch codes (for $t=1$ and $M=N$). Linear structure of batch codes can potentially be exploited in order to develop efficient retrieval algorithms. 
Therefore, linear batch codes are natural candidates for pratical applications, such as load balancing, CPIR and distributed storage. 
However, a lot of questions are 
remain open. We list some of them below. 
\begin{enumerate}
\item
Can the connection between linear batch codes and ECCs be extended to nonbinary codes? 
\item
Construct linear batch codes with better trade-offs between their parameters. 
\item
Construct linear batch codes suitable for distributed storage settings, in particular 
codes having sparse generator matrices. Obtain bounds on locality and repair bandwidth for such codes. 
\item
Do linear batch codes have as good parameters as their nonlinear counterparts do? 
\item
Develop efficient retrieval algorithms for batch codes.  
\end{enumerate}
Asnwering some of these questions could help in developing of new and more efficient batch codes, which can potentially be used in practical applications. 

\section*{Acknowledgement}

We thank Dominique Unruh for helpful discussions.


\end{document}